\documentclass[12pt, doublespace]{article}
\usepackage{graphicx}
\usepackage{epstopdf}
\title{ Inference for bounded parameters}

\author{D.A.S. Fraser and N. Reid\\ Department of Statistics, University of Toronto\\ Toronto Canada M5S 3G3\\
 A. Wong\\ Department of Mathematics and
Statistics,  York University\\ Toronto Canada M3J 1P3}

\begin{document} \maketitle
 
 \begin{abstract}
 The estimation of signal frequency count in the presence of background noise has had much discussion in the recent physics literature, and Mandelkern [1] brings the central issues to the statistical community, leading in turn to extensive discussion by statisticians.  The primary focus
however in [1] and  the accompanying discussion is on the construction of a confidence interval.   We argue that the likelihood function and $p$-value function provide a comprehensive presentation of the information available from the model and the data.  This is illustrated for Gaussian and  Poisson  models with lower bounds for the mean parameter.
 \end{abstract}

\section{INTRODUCTION}

\par Mandelkern [1] brings to the statistical community a seemingly simple
statistical problem that arises in high energy physics; see for example,
[2], [3].  The statistical model is quite elementary but
the related inference problem has substantial scientific presence: as Pekka Sinervo,
a coauthor of Abe et al. [2], [3] expresses, ``High energy physicists
have struggled with Bayesian and frequentist perspectives, with delays of
several years in certain experimental programmes hanging in the
balance".

The  problem discussed in [1] can be expressed simply.  A variable
$y$ follows a distribution with mean $\theta=b+\mu$, where $b>0$ is known, the shape of the 
distribution is known and the parameter $\mu\ge 0$.  The goal is to extract
the evidence concerning the parameter $\mu$, and in particular present the evidence on whether $\mu$
is zero or is greater than zero.  In the physics setting $y$ is often a count and is viewed as the sum
of a count of $y_1$  background events  and a count of $y_2$ events from a possible signal.  In [2] and [3], the signal records the presence of a possible top quark and the data come from the collider detector at Fermilab.  The background  count $y_1$
is modelled as Poisson$(b)$ and the count from the possible signal  as Poisson$(\mu)$.  
Following Mandelkern [1]
we  write $y \sim {\rm Poisson}(b+\mu)$ and let $\theta=b+\mu$ be the
Poisson mean with the restriction  $\theta\ge b$. There are additional aspects: for example the data are obtained as subsets of more complex counts, the background mean count $b$ is estimated and so on, but we concentrate on the simpler problem here.  We do however illustrate how the general case with $b$ estimated from 
antecedent Poisson counts can be treated within the general theory. 

The Poisson case involves a discrete distribution and this introduces some minor complications that that are best treated separately from the essential inference aspects.  Accordingly we  include a discussion of the continuous case
and for simplicity consider  the normal distribution for $y$ with mean $\theta=b+\mu$ and known standard deviation.

Much statistical literature  and most of the physics proposals cited by Mandelkern [1] are concerned with the construction of confidence bands for $\theta$ at some prescribed level of confidence.  It is our view that this leads to procedures that are essentially decision-theoretic: we ``accept'' parameter values within the confidence interval and ``reject'' parameter values outside the interval; a $1/0$ presentation.  This accept/reject theory evolved from Neyman and Pearson [4], later generalized as decision theory by Wald [5].  The decision theoretic approach dominated statistical theory until the mid 1950's, when Savage [6] promoted the personalistic Bayesian approach and Fisher [7] recommended an inference approach.  Both these approaches make essential use of the likelihood function: the Bayesian approach combines this with prior information, and the inference approach emphasizes the use the likelihood function and the observed significance  or $p$-value function.  The $p$-value function is constructed using the model and observed data, as we shall describe in more detail below.
One difficulty with the confidence interval approach arises from the presence of 
the lower bound $b$ for the parameter space;  if $y$ is small then the confidence interval can be partly or completely outside the permissible range $[b, \infty)$ for the parameter, making apparent
nonsense of an assertion of 95\% confidence.
Various proposals  have been put forward  to modify the confidence approach to overcome such difficulties; the most prominent being the unified approach of Feldman and Cousins [8].  These proposals seek an algorithm for placing a $1/0$ valuation on possible parameter values, in the framework of a prescribed confidence level.  By contrast the $p$-value function promoted here provides  essential evidence from the data concerning  the  value of the parameter; for some background see Fraser [9]. 

The discussants of Mandelkern [1] also focus on the confidence interval approach.  An exception is Gleser [10], who suggests the use of the ``likelihood function as a measure of evidence about the parameters of the model used to describe the data''; and Mandelkern [11] in his rejoinder concurs: ``it may be most appropriate to, at least in ambiguous cases, give up the notion of characterizing experimental uncertainty with a confidence interval ... and to present the likelihood function for this purpose.''  But also Abe et al [3] report the   $p$-value for the parameter value $\mu=0$; the $p$-value function extends this to all possible values of the parameter. The approach here recommends
the joint presentation of the likelihood function and the $p$-value function as the evidence from the data concerning the parameter. 

In Section 2 we record some discussion of the unified approach and its variants, and also record various anomalies associated with their use.

In Section 3 we expand on Mandelkern's comment and discuss what we call an inferential approach. This records the observed likelihood function and the observed $p$-value function.  We feel that these present the full statistical evidence concerning the parameter, and  in turn allow appropriate individual judgments to be made concerning the parameter.  An experiment reported in [3] is analysed using the Poisson model with background, first for known background and then allowing Poisson variation in the background.  

\section{The unified approach and variants}
The construction of a confidence interval is often based on the theory of optimal testing, and this can lead to rather anomalous behavior.   An optimality criterion typically involves averaging over the sample space, and in many situations there are what Fisher [7, 12] called `recognizable subsets' of the sample space, subsets  that appropriately partition the sample space.  In this setting the use of overall optimality can mean that intervals are constructed which effectively trade performance in a single instance for average performance in a series of instances, most of which may have recognizably different features.    In
extreme cases this can give a confidence interval that is empty or a
confidence interval that is the full range for the parameter: in such
cases the overt confidence is clearly zero or 100\% in contradiction to
the prescribed or targetted confidence.  For some recent discussions
with examples, see Fraser [9] where the optimality criteria are shown to lead to decisions that are contrary to the available evidence; see also Cox [13] on the general appropriateness of optimality criteria.

The conventional  intervals applied to examples with  a bounded parameter space also can lead to anomalous confidence intervals. 
Thus an optimum confidence interval derived for the unrestricted case may well
lap into the inappropriate region $\theta < b$, this being the key issue
in the Poisson case and mentioned for the continuous case in Mandelkern [1].
  
Various proposed modifications to the typical central confidence interval are discussed in Mandelkern [1].  Assume we have a scalar variable $y$ with a continuous density $f(y;\theta)$ and with a distribution function $F(y;\theta)$ that is stochastically increasing in $\theta$.  Denote by $y_L(\theta)$ and $y_U(\theta)$ the $\gamma$ and 95\% + $\gamma$
quantiles of $F(y;\theta)$; these form a 95\% confidence interval.
Now let $\gamma=\gamma(\theta)$ vary with $\theta$ but be
restricted to the interval $(0, 5\%)$.  The confidence
belt in the $y\times\theta$-space is the set union of 
the acceptance regions $(y_L(\theta), y_U(\theta)) \times
\{\theta\}$;  and  the $y$-section
of the two dimensional confidence belt is a 95\% confidence region 
and under moderate regularity will have the form  
$(\theta_L(y), \theta_U(y))$. A reasonable  objective  is to have these sets stay
within the acceptable range $[b,\infty)$ by some natural-seeming choice
of the adjustment function $\gamma(\theta)$.

The likelihood ratio is used as one basis for deciding which points 
are to go 
into 
the acceptance interval $(y_L(\theta), y_U(\theta))$
and thus for determining $\gamma(\theta)$.
Then to form the acceptance 
interval the points are ordered from the largest using
the ratio 
\begin{equation} R=\frac{L(\theta;y)}{L(\tilde\theta;y)}
\end{equation} 
where $L(\theta;y)=f(y;\theta)$ and 
$\tilde\theta=\tilde\theta(y)$ is a
reference parameter value to be used with $y$.  The Unified Approach of
Feldman \& Cousins [8] takes $\tilde\theta=\hat\theta(y)$ to be the
maximum likelihood estimate of $\theta$ under the restriction $\theta\ge b$; for example 
in the Normal$(\theta,1)$ case, we have  $\hat\theta=\max(b,y)$.  The New
Ordering approach of Giunti [14] takes $\tilde\theta$ to be a Bayesian
expected value for $\theta$.  Using a somewhat different starting point Mandelkern \& Schultz [15] obtain likelihood
from the distribution of $\hat\theta(y)$, which is a marginalisation
from the distribution of $y$ itself. For the normal case this $\hat\theta$
does not depend on $y$ for $y < b$ and not surprisingly the confidence
intervals obtained by this approach are found not to depend on $y$ for
$y < b$; the resulting intervals had been considered earlier by Ciampolillo [16].
The use of these optimizing or ordering criteria can have rather strange  effects. For,
as noted, the criteria involve shifting the distribution bound to the left for low parameter values so that the $2.5\%$
tail probabilities on the left and the right are changed to have less on the left and more on the right;
this has the effect for small data  values of shifting the confidence intervals to the right, away from the excluded parameter value range. The disturbing result however is that the lower confidence bound is no longer
a $2.5\%$ bound but something larger and perhaps undefined. And the upper confidence 
bound is no longer a $97.5\%$ bound but something larger and perhaps undefined.  Thus the individual bounds of the confidence interval do not have the direct  statistical meaning that one would reasonably impute to them; this is particularly serious and disturbing in a context where the lower bound is directly addressing the issue of whether or not $\mu$  is
equal to zero.
These approaches seem to seek a single construction that combines the merits of one-sided and two-sided confidence intervals.  In a sense this is treating both $b$ and $\theta$ as parameters and having the same construction provide conclusions about both of them.  The inferential approach of the next section emphasizes the evidence in the data about the single parameter $\theta$, with $b$ fixed.  The extension to the case of estimated background  illustrated in Section 4 emphasizes the evidence in the data about $\theta$, in the presence of a nuisance parameter.


For the Poisson problem described in the introduction,  Roe \&
Woodroofe [17]  propose the use of certain conditional probabilities as
the basis for the confidence belt construction following Feldman \& Cousins [8].  Such conditioning had been proposed earlier for upper limits by Zech [18].  
Roe and Woodroofe [17] recommended the use of the
conditional distribution of $y$ given $y_1\le y^0$, say
$g(y;\mu)=f(y|y_1\le y^0;\mu)$ as recorded (4) in Mandelkern
[1].  But the variable $y_1$ is not an observable variable and hence not ancillary in the usual sense, and the proposed conditioning does not generate a partition of the sample space. This was noted in Woodroofe \& Wang [19] and in Cousins [20], and a Bayesian
approach was proposed in Roe and Woodroofe [21]. 
Thus the nominal conditional distribution does not satisfy the standard conditions for validity in describing conditional frequencies given observed information.
Also not surprisingly, as noted by Mandelkern [1] and Cousins [20],
there is a related 
undercoverage which can be severe
for the nominal confidence intervals constructed.

\section{The statistical evidence: the likelihood and $p$-value functions}


Consider first a sample $\underline y = (y_1, \dots , y_n)$ from the
Normal $(\theta, \sigma_0^2)$ distribution with $\sigma_0^2$  known.  The likelihood function is proportional to the density for the sample mean at the observed value $\bar y^0$, and is examined as a function of the unknown $\theta$:
\begin{equation}
L(\theta)=c\phi(n^{1/2}(\bar y^0-\theta)/\sigma_0),
\end{equation} 
where $\phi$ is the standard normal density. 
The $p$-value function is the probability  that the sample mean is less than or equal to  the observed
$\bar y^0$:
\begin{equation}
p(\theta) = \Phi(n^{1/2}(\bar y^0-\theta)/\sigma_0),
\end{equation} 
where  $\Phi$ is the standard
normal distribution function.    The $p$-value function uses the known sampling distribution of $\bar y$, and records the percentile position of the observed data in the distribution having
parameter value $\theta$.   The more conventional interpretation of the $p$-value as ``the probability of observing a result as or more extreme, under the model" is obtained as 1 minus the $p$-value function when the data is in the right tail of the distribution.  Two-tailed $p$-values can also be obtained if desired.
As a function of $\bar y$, $p(\theta)$ is uniformly distributed on $(0,1)$ under
the assumed model.  This  ``repeated sampling'' property of
the $p$-value is the analogue of coverage of a confidence interval.   

This discussion extends directly to any location model $f(y-\theta)$ for $y$.  The likelihood function is 
\begin{equation}
L(\theta)=L(\theta;y_1^0, \dots , y_n^0)=c(\underline y^0)\prod_{i=1}^n f(y_i^0-\theta) .
\end{equation}
And the $p$-value function, using the sampling distribution of $\bar y$ conditional on the observed sample configuration $a^0=(y_1^0-\bar y^0, \dots, y_n^0-\bar y^0)$, is
\begin{equation}
p(\theta)=\int_{-\infty}^{y^0}  f(\bar y \mid a^0)d\bar y ;  
\end{equation}
in the special Gaussian case $\bar y$ is independent of $a$.
This raises essentially no new problems beyond the computation of the integral. For this
location model it can be shown that the $p$-value function is identical to the integral of the likelihood function, so that
\begin{equation}
p(\theta)= \int_\theta^\infty L(\nu)d\nu /\int_{-\infty}^\infty L(\nu)d\nu;
\end{equation}
thus the $p$-value function is identical to the posterior survivor function (one minus the posterior distribution function) using the flat prior $\pi(\theta)d\theta=d\theta$.
 
The location form of the model provides a procedure for simplifying the data vector $(y_1, \dots , y_n)$ to a scalar summary, $\bar y$, by conditioning.  As the distribution of the sample configuration is free of $\theta$, no information is lost by this conditioning.  The use of $\bar y$ as the one-dimensional variable is not essential; the same result is obtained using the maximum likelihood estimate, or in fact any location respecting estimator of $\theta$, together with a notational change in the expression for $a$.   In the methods for more general models this argument is applied using approximate conditioning and  reexpression of the parameter to location form.   

We now return to the Poisson $(\theta)$ with
$\theta=b+\mu$ where $b$ is known and $\mu\ge 0$.  The Poisson case is simpler, in that the model specifies a one-dimensional variable, $y$, and a one-dimensional parameter $\theta$, so no dimension reduction is needed.  The likelihood
function from an observed count $y^0$ is 
\begin{equation}
 L(\theta)=c\theta^{y^0} e^{-\theta} 
 \end{equation}
  where
$\theta=b+\mu$.  This can be plotted as a function of 
$\mu$ for $\mu$ in $(-b, \infty)$: for $\mu$ in $[0, \infty)$ it describes the probability at the observed data point under the assumed model; for $\mu$ in $[-b, 0)$ it serves as a diagnostic
concerning $b$, suggesting that either the model or the computation of $b$ is not correct.  The $p$-value 
function at $y^0$ is given by the interval 
\begin{equation}
p(\theta)= (F^-(y^0; \theta),F(y^0; \theta))
\end{equation}
of  numerical values,
where $F(y; \theta)$ is the
Poisson$(\theta)$ distribution function and $F^-(y; \theta)$ is
the probability up to, but not including, $y$ and is given by
$F(y-1; \theta)$.  We use an interval of $p$-values in accord with the discreteness in the problem; a compromise is to plot the so-called {\em mid p-value}, which is
$F^-(y^0; \theta)+(1/2)f(y^0; \theta)$.
In our approach an observed $y^0$ leads  to a continuum of
numerical $p$-values for each $\theta$ being assessed.  This proposal acknowledges the discreteness explicitly and yet does maintain the repeated sampling property of the $p$-value function, that it have a uniform distribution on $(0,1)$. 
Other aspects of the discreteness problem are addressed in Brown et al. [22] and Baker [23].

As a simple example consider
$b=2$ with data $y^0=3$. The likelihood and $p$-value functions are
recorded in Figure 1.  
The likelihood for $\mu$ is easily understood, and particularly useful when combining data.  The interpretation of a $p$-value for  given data value is exactly analogous to the percentile score on, for example, a standardized test:  it expresses the percentile position of the data point relative to the parameter.  For the null condition $\theta = 2$ or $\mu = 0$ the $p$-value interval for the data $y^0=3$ is $(0.677 0.857)$; a fairly central and broad range.  

If $y^0=0$, and $b>0$, then $p(\mu)=(0,\exp\{-(b+\mu)\})$.  This emphasizes the
lack of information in the data about $\mu$, and this lack is most striking when $\mu =0$ and $b$ is very small. For larger $b$, the observed value of $0$ will be further in the left tail of the $\mu=0$ distribution.

In Abe et al. [3] after preliminary simplification from their Table 1
we have $b = 6.7$ with $y^0=27$. The likelihood function and $p$-value
functions are plotted in Figure 2. For the null condition $\theta = 6.7$ or $\mu = 0$ the data is in the extreme right tail and the upper and lower $p$-values are essentially 1.  The actual values are 
$(1-3\times 10^{-8}, 1- 10^{-8})$ thus offering very strong evidence that $\mu>0$.
\begin{figure}
\begin{center}
\caption{The likelihood function (top) and $p$-value function (bottom) for the Poisson model, with $b=2$ and $y^0=3$.  For $\mu=0$ the $p$-value interval is $(0.677, 0.857)$.}
\centerline{\includegraphics[width=3.5in,height=5.5in]{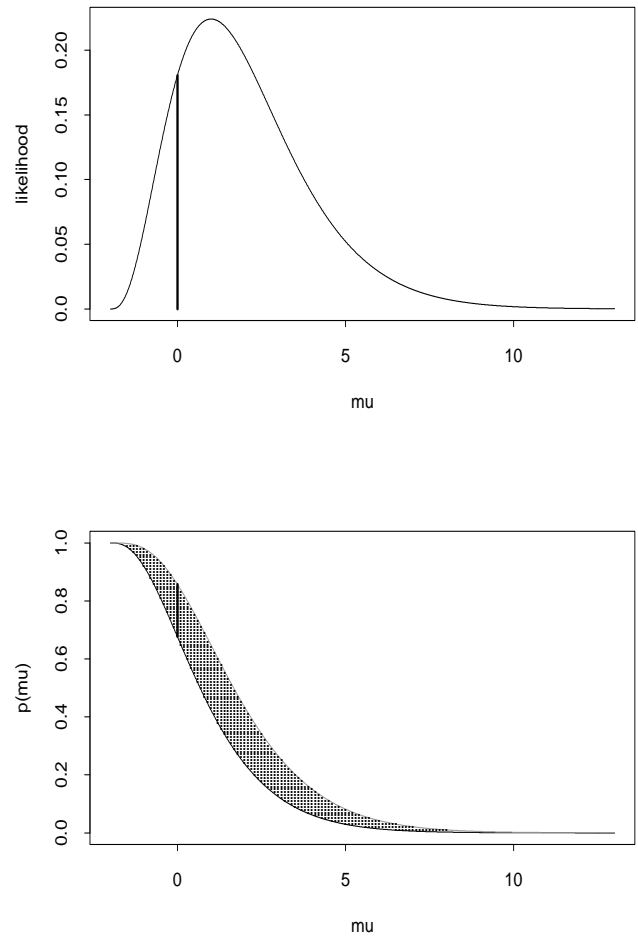}}
\end{center}
\end{figure}

\begin{figure}
\begin{center}
\caption{The likelihood function (top) and $p$-value function (bottom) for the Poisson model, with $b=6.7$ and $y^0=27$.  For $\mu=0$ the upper and lower $p$-values are essentially 1.}
\centerline{\includegraphics[width=3.5in,height=5.5in]{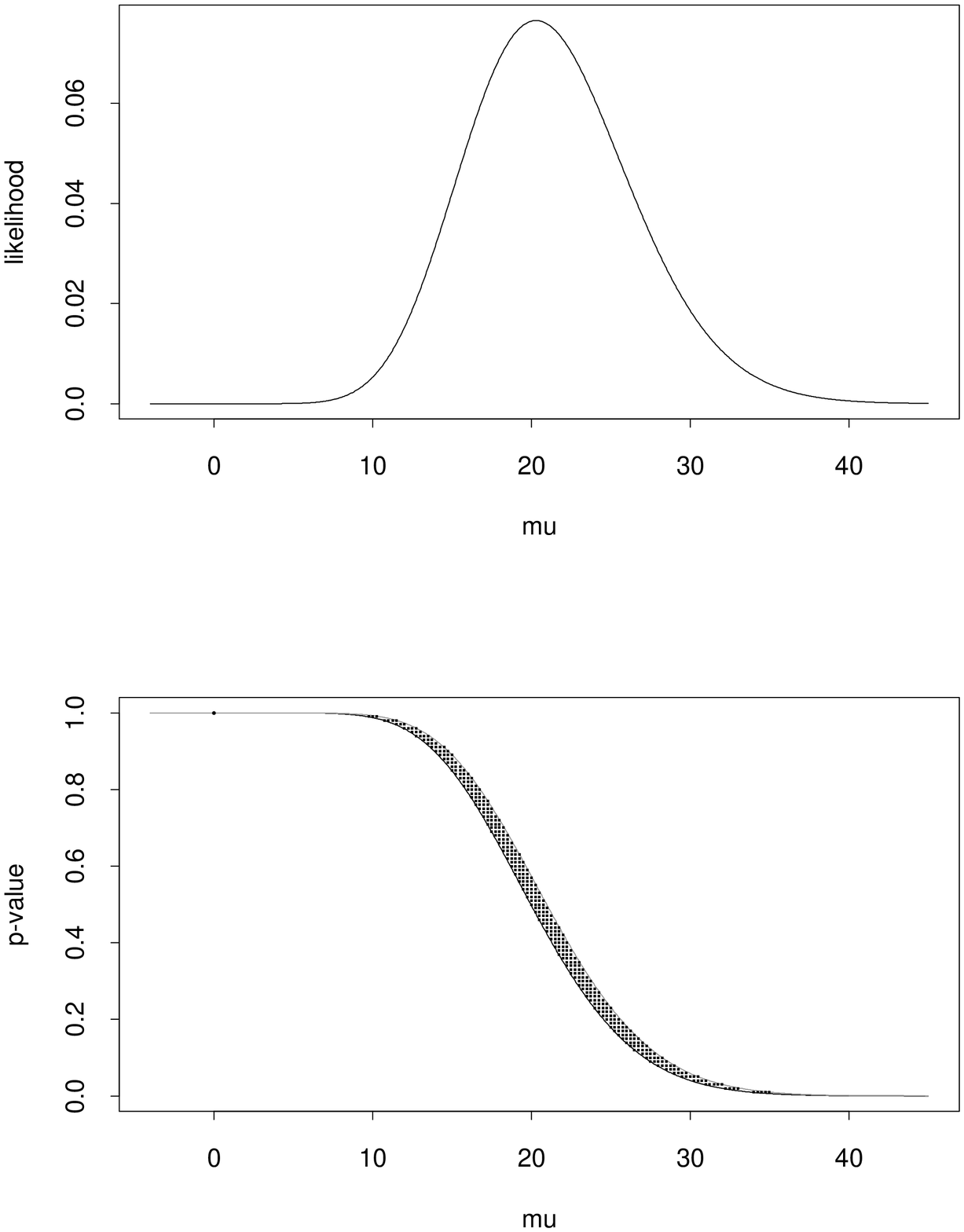}}
\end{center}
\end{figure}

Figure 3 shows the corresponding likelihood and $p$-value plot for a Gaussian model, with $\mu=b+\theta$, where $n=1$, $\sigma=0.5$ and  $y=1.8705$ with $b$ taken to be 1.4142.     

The Gaussian case is not as far removed from the Poisson as might be thought at first.  If $y \sim $Poisson$(\theta)$, then $\surd y$ is approximately distributed as Gaussian with mean $\surd\theta$ and standard deviation $1/2$, at least for large $\theta$.  For comparison with the first example and Figure 1, the $p$-value interval for testing $\mu=0$  computed using the normal approximation with continuity correction, i.e. evaluated at $\surd(y^0\pm  0.5)$,  is (0.631, 0.819).  

\begin{figure}
\begin{center}
\caption{The likelihood function (top) and $p$-value function (bottom) for the normal approximation to the Poisson  model, after square root transformation with data as in Figure 1.}
\centerline{\includegraphics[width=3.5in,height=5.5in]{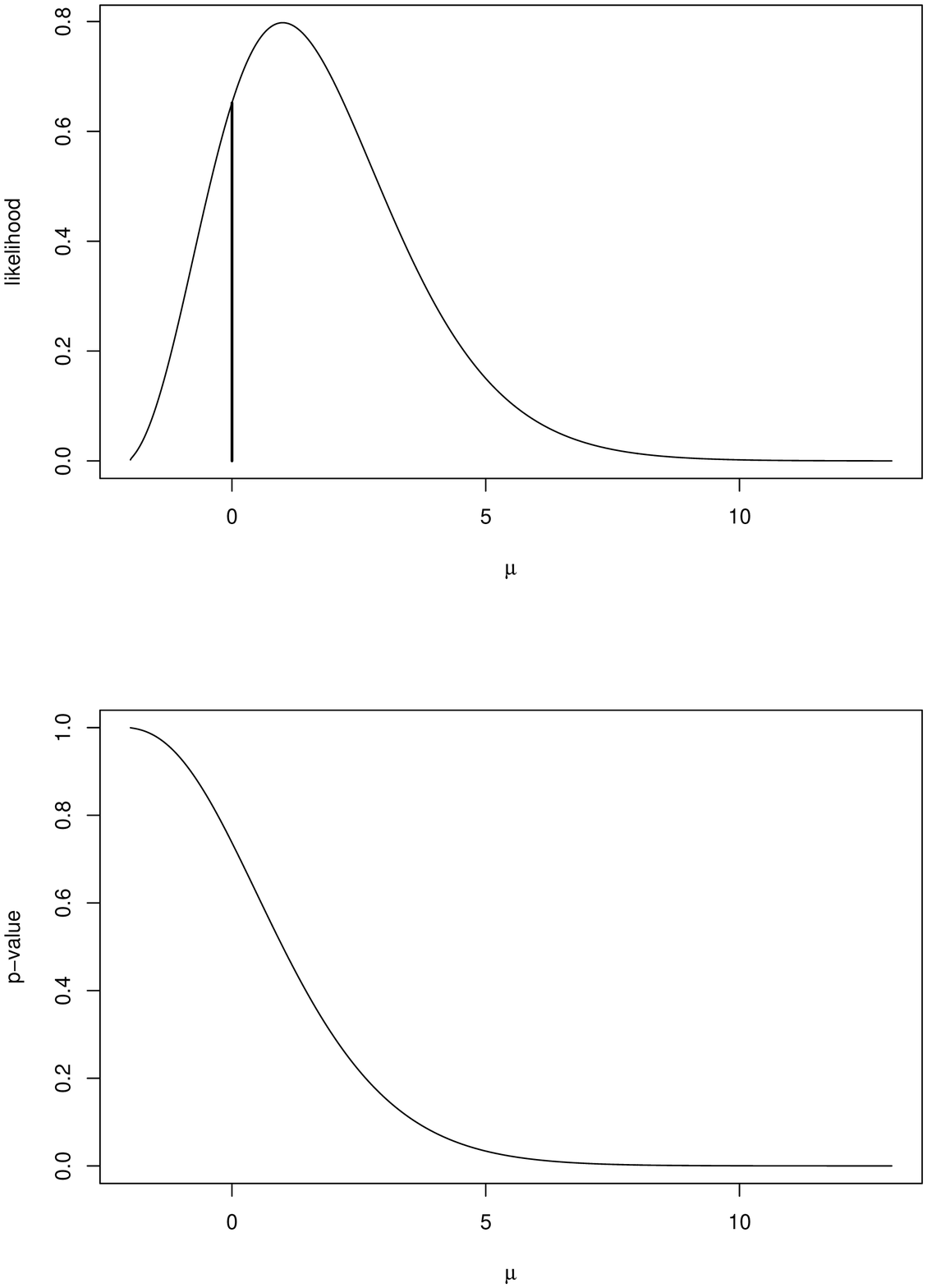}}
\end{center}
\end{figure}

It is possible to use recently developed work in likelihood asymptotics in the Poisson model with estimated background.  We suppose that the background mean count $\beta$ is an unknown parameter estimated by $b$.  To reflect the precision in this
estimate, we write
\begin{equation}
b=y_1/k
\end{equation}
where $y_1$ follows a Poisson distribution with mean $k\beta$ and hence variance
$k\beta$.  A value for the standard error of $b$, say $\sigma_b$ determines a value
for $k$ as $k=b/\sigma_b^2$.  In [3] the estimated standard error from Table II is 2.1, with $b=6.7$.  The resulting $p$-value function is plotted in Figure 4, where it is compared with the mid $p$-value function assuming the background is known.  The value of the new $p$-value function at $\mu=0$ is $1-2.6\times 10^{-5}$.

\begin{figure}
\begin{center}
\caption{The $p$-value function using the third order approximation developed from [25],  allowing for estimation errors in the background signal, compared with the mid $p$-value assuming the background is known.}
\centerline{\includegraphics[width=5.5in,height=5.5in]{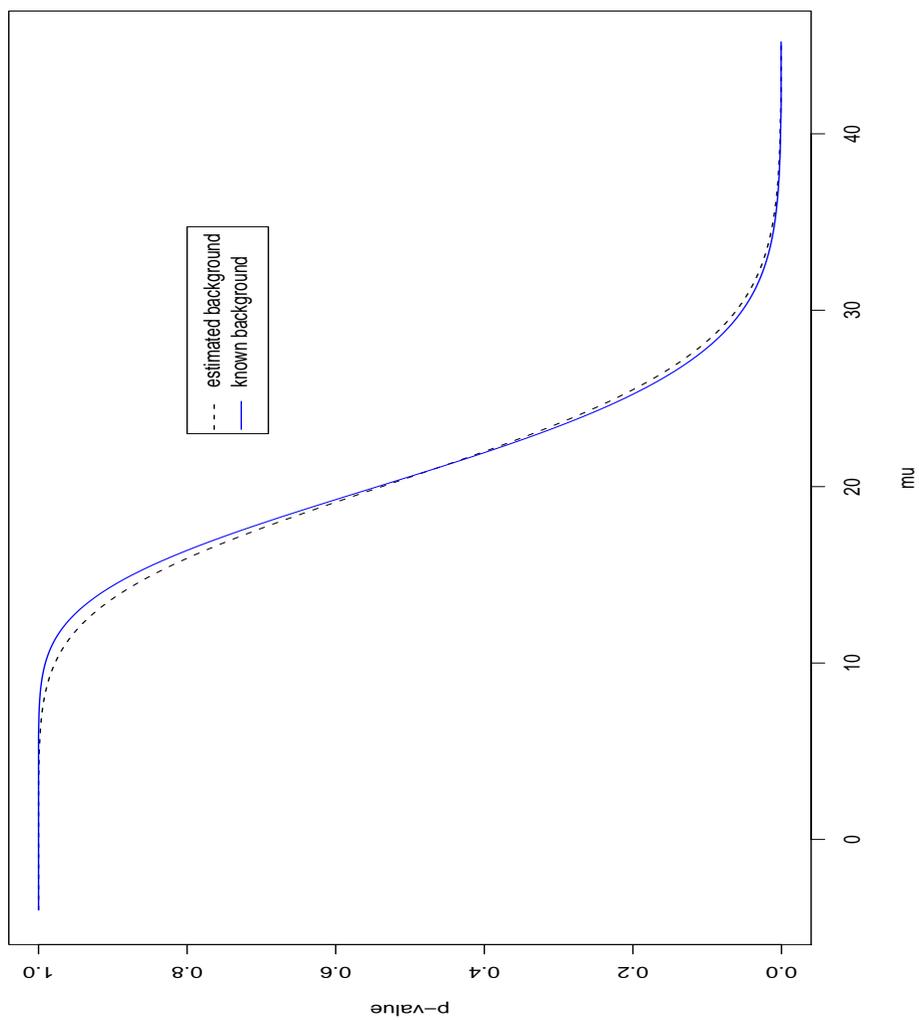}}
\end{center}
\end{figure}

\section{Discussion}

The $p$-value function, evaluated at a particular value $\theta_0$, gives the percentile position of the observed data relative to the model with that parameter value $\theta_0$. Our view is that the $p$-value provides the key scientific evidence in the data relative to the assumed model. In contrast a fixed level confidence approach provides a much more limited statement that the parameter is or is not contained in a given interval.   An improvement to the confidence approach would be the reporting of confidence limits at a continuum of confidence levels, which is mathematically close to the $p$-value function approach.   One can  use the $p$-value function to construct a confidence interval at level $1-\alpha$, by finding the parameter values for which the $p$-value equals, say, $1-\alpha/2$ and $\alpha/2$.  However our definition of  the $p$-value function is intrinsically one-sided, as seems  more appropriate for the physical context of detecting a signal. 

It is important to know how the inferential approach promoted here generalizes to more complex models.  Most realistic models will have a parameter $\theta$ of dimension $d$, say.  For this setting we might be interested in a scalar component $\psi(\theta)$ and could then want the $p$-value function for $\psi$.  If more than one component of $\theta$ is of particular interest each could be examined  in turn.  The essential simplification available for this setting from recently developed likelihood theory  is  that to a high order of approximation there is a conditional model that behaves like a location model for $\psi$ with  a related scalar variable that  measures this parameter.  Approximations to the corresponding observed likelihood function are given in Fraser [24] and approximations to the $p$-value function are given in Fraser, Reid and Wu [25].  These evolved from a closely related approach based on ancillarity due to Barndorff-Nielsen  which  is summarized in Barndorff-Nielsen and Cox [26].  The approach as described in [25] requires that $y$ follow a continuous distribution in general models.  Work in progress with A.C. Davison extends this approach to the discrete setting, and this work was used to derive the results summarized in Figure 4.  

The statistical literature summarizing higher order likelihood asymptotics is still fairly specialized, but some review or book length treatments are available in Reid [27], Severini [28], Skovgaard [29] and Barndorff-Nielsen and Cox [26].

\bigskip
\noindent{\bf Acknowledgements}

The authors wish to thank the referees for many very helpful comments that assisted in the revision of an earlier version.  The research was partially supported by the Natural Sciences and Engineering Research Council of Canada.

\newpage

\par
[1] M. Mandelkern, 
{\it Statist. Sci.} {\bf 17} 149 (2002).

[2] F. Abe et al.,  {\it Phys. Rev. Lett.} {\bf 73(2)} 
225 (1994).

[3] F. Abe et al.,  {\it Phys. Rev. Lett.} {\bf 74(14)} 2626 (1995).

[4] J. Neyman and E. S. Pearson, {\it Phil. Trans. R. Soc.} A {\bf 239} 289 (1933).

[5] A.Wald, {\it Statistical Decision Functions.} New York: Wiley (1950).

[6] L. J. Savage, {\it The Foundations of Statistics.} New York: Wiley (1954).

[7] R.A. Fisher, {\it Statistical Methods and Scientific Inference.}
Edinburgh: Oliver and Boyd (1956).

[8] G.J. Feldman and R.D. Cousins, {\it Phys. Rev.} D {\bf
57} 3873 (1998).

[9] D.A.S. Fraser, {\it Statist. Sci.}, to appear (2003).

[10] L.J.  Gleser, {\it Statist. Sci. }{\bf 17} 161 (2002).

[11] M. Mandelkern, {\it Statist. Sci. }{\bf 17} 171 (2002).

[12] R.A. Fisher, {\it Proc. Roy. Soc. } {A} {\bf 144}, 285 (1934).

[13] D.R. Cox, {\it Ann. Math. Statist.} {\bf 29}, 357 (1958).

[14] C. Giunti,
{\it Phys. Rev.} D {\bf 59} 113000 (1999).

[15] M. Mandelkern and J. Schultz, {\it J. Math. Phys.} {\bf 41} 5701 (2000).

[16] S. Ciampolillo, {\it Il Nuovo Cimento} {\bf 111}, 1415 (1998).

[17] B. P. Roe and M.B. Woodroofe, {\it Phys. Rev.} D {\bf
60} 053009 (1999).

[18] G. Zech, {\it Nuclear Instruments and Methods} {\bf A277}, 608 (1989).

[19] M.B. Woodroofe and H. Wang, {\it Ann. Statist.} {\bf 28} 1561 (2000).

[20] R.D. Cousins, {\it Phys. Rev. D} {\bf 62} 098301 (2000).

[21] B.P. Roe and M.B. Woodroofe, {\it Phys. Rev.} D {\bf 63} 013009 (2001).

[22] L.T. Brown, T.T. Cai and A. DasGupta, {\it Statist. Sci.} {\bf 16}, 101, (2001).

[23] L. Baker, {\it Amer. Statist.} {\bf 56} 85, (2002).

[24] D.A.S. Fraser, {\it Biometrika} {\bf 90} 327 (2003).

[25] D.A.S. Fraser, N. Reid, J. Wu, {\it Biometrika} {\bf 86} 249 (1999).

[26] O.E. Barndorff-Nielsen and D.R. Cox. {\it Inference and Asymptotics.} , Boca-Raton: Chapman \& Hall/CRC.

[27] N. Reid, {\it Ann. Statist.}, to appear (2003). \\ available at {\tt www.utstat.utoronto.ca/reid/research}.

[28] T.A. Severini. {\it Likelihood Methods in Statistics}  Oxford: Oxford University Press (2000).

[29] I.M. Skovgaard, {\it Scand. J. Statist. } {\bf  28}, 3--32, (2001).

\end{document}